\documentclass[runningheads]{llncs}
\usepackage{graphicx}
\usepackage{hyperref}
\usepackage{xcolor}

\usepackage{amsmath}
\usepackage{amssymb}
\usepackage{multirow}
\usepackage{bbm}
\usepackage{algorithm}
\usepackage[noend]{algpseudocode}
\begin{document}
\title{Self-Supervised k-Space Regularization \\ for Motion-Resolved Abdominal MRI Using Neural Implicit k-Space Representations}
\titlerunning{PISCO: Self-Supervised k-Space Regularization}
\author{
Veronika Spieker \inst{1,2,3*} 
\and Hannah Eichhorn \inst{1,2} 
\and Jonathan K. Stelter \inst{2} 
\and \\ Wenqi Huang \inst{2} 
\and Rickmer F. Braren  \inst{2} 
\and Daniel Rückert \inst{2,4} 
\and  Francisco \\ Sahli Costabal \inst{3,5} 
\and Kerstin Hammernik  \inst{2} 
\and  Claudia Prieto  \inst{3,5,6} 
\and \\ Dimitrios C. Karampinos  \inst{2} 
\and Julia A. Schnabel \inst{1,2,6}
}
\authorrunning{V. Spieker et al.}
%
\institute{
 Helmholtz Center Munich, Germany,  \\
\and Technical University of Munich, Germany\\
\and Millenium Institute for Intelligent Healthcare Engineering, Chile \\
\and Imperial College London, United Kingdom \\
\and Pontificia Universidad Católica de Chile, Chile\\
\and King’s College London, United Kingdom \\
*\email{v.spieker@tum.de}
} 

\maketitle              
\begin{abstract}
Neural implicit k-space representations have shown promising results for dynamic MRI at high temporal resolutions. Yet, their exclusive training in k-space limits the application of common image regularization methods to improve the final reconstruction. In this work, we introduce the concept of parallel imaging-inspired self-consistency (PISCO), which we incorporate as novel self-supervised k-space regularization enforcing a consistent neighborhood relationship. At no additional data cost, the proposed regularization significantly improves neural implicit k-space reconstructions on simulated data. Abdominal in-vivo reconstructions using PISCO result in enhanced spatio-temporal image quality compared to state-of-the-art methods. Code is available at \url{https://github.com/vjspi/PISCO-NIK}.

\end{abstract}

\section{Introduction}

Motion, i.e. induced by respiration in the abdomen, results in non-negligible artefacts in magnetic resonance imaging (MRI). Therefore, motion-sensitive applications such as radiation therapy planning \cite{Stemkens_2018} or free-breathing high-resolution diagnostic imaging \cite{Zaitsev_2015} rely on dynamic MRI reconstruction techniques. These techniques reconstruct temporally resolved images by binning the data into different respiratory motion states (MS).
However, reconstructing multiple MS inherently results in a reduction of data per temporal image, leading to undersampling artifacts due to violation of the Nyquist theorem. Common approaches address this by applying regularization across the temporal dimension to exploit redundancies \cite{Feng_2016,Terpstra_2023}. Yet, the total number of MS is limited, resulting in motion blurring due to low temporal resolution.

Recently, neural implicit k-space representations (NIK) have shown promising results for blurring-free motion-resolved reconstructions 
\cite{Spieker_2024,Huang_2023}. Based on the acquired k-space trajectory and a surrogate signal for the current motion state, a multi-layer perceptron (MLP) is trained to predict the k-space signal for a given spatio-temporal input. At inference, any coordinate can be sampled, allowing for flexible spatial sampling patterns at high temporal resolutions. Exclusive training in k-space eliminates the need for costly domain transforms such as non-uniform Fast Fourier Transformations (NUFFT) within each iteration.

Similar to other accelerated reconstruction methods \cite{Akcakaya_2019,Griswold_2002}, NIK learns to fill missing data points within k-space. Yet, without any kind of regularization, the reconstruction model may be prone to overfitting and noisy reconstructions. General learning-based MRI reconstruction methods counteract overfitting by including a regularization, usually enforced on the estimated image \cite{Ahmad_2020,Hammernik_2023,Jafari_2023}. However, image-based regularization methods are impractical for NIK, as they necessitate sampling of the entire k-space within each training iteration to obtain the image that requires regularization. While k-space-based regularization would be effective, translation of image-based constraints to k-space is not trivial. 

Exploring the parallel imaging concept of Generalized Autocalibrating Partially Parallel Acquisitions (GRAPPA) \cite{Griswold_2002} reveals a potential spatial neighborhood relationship within k-space itself. This relationship needs to be estimated on a fully-sampled calibration set to be subsequently applied to undersampled k-space regions. Learning-based reconstruction methods already utilize this neighborhood relationship for k-space refinement \cite{Ryu_2021,Spieker_2023}. However, similar to GRAPPA, they first require explicit determination of the k-space relationship
This is impractical for motion-resolved imaging, since calibration needs to be acquired and conducted for every MS. 

In this work, we reformulate the concept of k-space 
relationship independent of calibration data. We exploit the inherent global k-space relationship without the need for explicit determination. Our contributions are three-fold:
\begin{enumerate}
    \item We introduce PISCO: a novel Parallel Imaging-inspired Self-Consistency for self-supervised k-space refinement that operates independently of any additional training or calibration data. 
    \item Incorporating the concept of PISCO, we present the first k-space-based regularization loss for neural implicit k-space representations (NIK).
    \item Based on a realistic motion simulation as well as in-vivo data, we quantitatively and qualitatively demonstrate the potential of PISCO for motion-resolved abdominal MR reconstruction using NIK.
\end{enumerate}

\section{Methods}
\subsection{k-Space Interpolation using GRAPPA}
\label{sec:methods_grappa}
In MRI, a k-space signal $y=\{y_i\in \mathbbm{C}^{N_c}|i=1,...,N_x \cdot N_y\}$ is acquired with $N_c$ coils at spatial k-space coordinates $k=\{k_i \in \mathbbm{R}^{2}|i=1,...,N_x \cdot N_y\}$ (no time dimension for simplicity). The desired image ${x} \in \mathbbm{C}^{N_x \cdot N_y}$
is then reconstructed by solving the inverse problem $y = \mathrm{A} x $, where the forward operator $ \mathrm{A} = \mathcal{F} \mathbf{S} $ consists of the coil sensitivity maps $\mathbf{S} \in \mathbbm{C}^{N_x \cdot N_y}$ and Fourier transform $\mathcal{F}$. For acceleration purposes the acquired k-space is often undersampled, i.e., $y$ is acquired at fewer $k_i$ than required to fulfill the Nyquist theorem. To avoid undersampling artefacts, parallel imaging methods such as GRAPPA \cite{Griswold_2002} leverage the multi-coil nature of MRI data to derive missing values from its neighboring k-space values.

\begin{figure}[t]
\includegraphics[width=\textwidth]{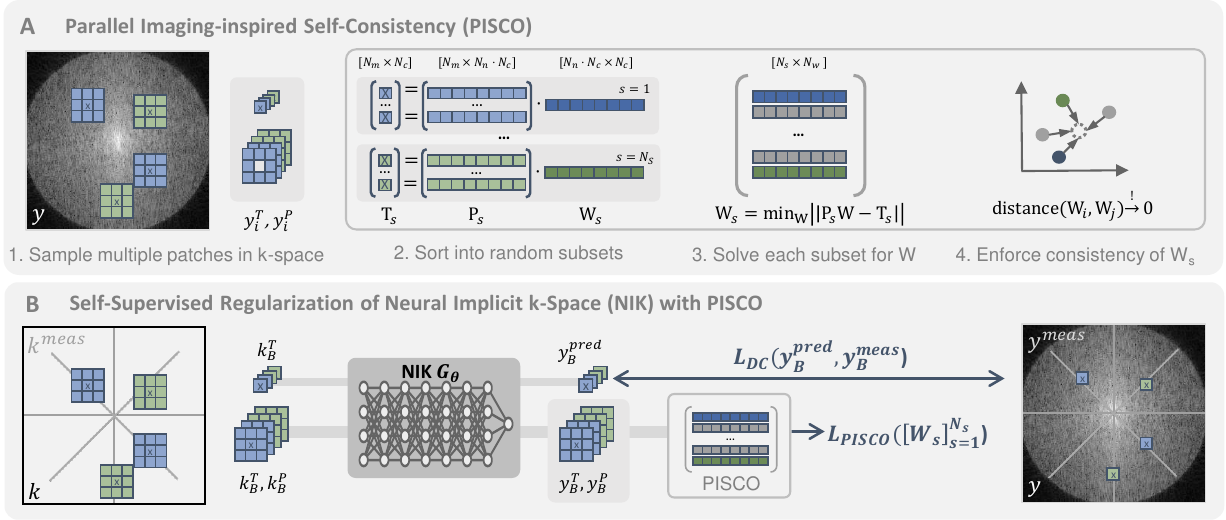}
\caption{Overview. A: Multiple pairs of targets and surrounding neighbors $y_i^T, y_i^P$ are sampled and randomly sorted into subsets and solved for the linear relationship $W_s$ (Eq. \ref{eq:PISCO_weightsolving}). PISCO aims to minimize the distance between all $\mathrm{W_s}$. For simplicity, the coil dimension $N_c$ is not visualized, but included in matrix dimensions. B: PISCO for regularization of NIK. Any points can be sampled, but $L_{DC}$ can only be compared to measured k-space $y_{meas}$ (gray lines). $L_{PISCO}$ refines independent of $y_{meas}$.} 
\label{fig:overview}\end{figure}

For GRAPPA, a patch of $N_n$ neighboring coordinates is sampled around a missing target $k^T_{i} \in \mathbbm{R}^{2}$, resulting in a coordinate patch $k^P_{i} \in \mathbbm{R}^{N_n \cdot {2}}$. The target signal value $y^T_i \in \mathbbm{C}^{N_c}$ is then obtained by linearly combining the neighboring signal values $y^P_{i} \in \mathbbm{C}^{N_n \cdot N_c}$ (Fig. \ref{fig:overview}A). Stacking $N_m$ pairs of targets and patches, the linear combination can be written as $ \mathrm{T} = \mathrm{P}\mathrm{W}$, where $\mathrm{T} = [y^T_{1}, ..., y^T_{N_m}] \in \mathbbm{C}^{[N_m \times N_c]}$, $\mathrm{P} = [y^P_{1}, ..., y^P_{N_m}] \in \mathbbm{C}^{[N_m\times N_n \cdot N_c]}$ and $\mathrm{W}\in \mathbbm{C}^{[N_n \cdot N_c\times N_c]}$ is the global weight matrix with a total of $N_w = N_n \cdot N_c \cdot N_c$ weights.
To determine $\mathrm{W}$, GRAPPA splits the signal space $y$ into a fully sampled auto-calibration signal $y_{\mathrm{ACS}}$ and the remaining undersampled k-space $y_{\mathrm{US}}$. Within the fully sampled $y_{\mathrm{ACS}}$, pairs of $\mathrm{T_{\mathrm{ACS}}}$ and $\mathrm{P_{\mathrm{ACS}}}$ can be used to solve the least-squares problem: 

\begin{equation}
   \mathrm{W_{ACS}} = \arg \min_\mathrm{W}  \lVert \mathrm{P_{ACS} }\mathrm{W} - \mathrm{T_{ACS}} \rVert_2^2 +  \alpha \lVert \mathrm{W} \rVert_2^2 \text{\quad s.t. } \mathrm{T}_{ACS}, \mathrm{P}_{ACS} \subseteq y_{ACS}
\end{equation}
where $\lVert \cdot \rVert_2^2$ applies the L2-norm element-wise and $\alpha$ weighs the Tikhonov regularization. Subsequently, undersampled points $y^T_{US}$ are derived using the calibrated weights to compute $ y^T_{US} = \mathrm{W_{ACS}}  \cdot y^P_{US}$, leading to improved reconstructions \cite{Griswold_2002}.

\subsection{From GRAPPA to PISCO}
\label{sec:methods_grappa2pisco}
Acquiring a fully sampled region $y_{ACS}$ to calibrate a weight set $\mathrm{W} $ is not always feasible, particularly in dynamic imaging. Hence, we reframe GRAPPA's concept of global spatial k-space relationship to a calibration-free condition based on the following assumption: 
If a weight set $\mathrm{W}_{\mathrm{ACS}}$ calibrated on $y_{\mathrm{ACS}} \in y $ models the linear relationship for the whole k-space $y$, then a weight set $\mathrm{W}_\mathrm{s}$ derived from a random subset $y_\mathrm{s} \in y $ should result in the same linear relationship. Consequently, weight sets $\mathrm{W}_\mathrm{s}$ determined from multiple random subsets $\{y_\mathrm{s} \in y| s= 1,...,N_s\}$ are expected to converge to the same solution, i.e. $\mathrm{W}_i = \mathrm{W}_j, \forall i \neq j$. Thus, we propose to refine an interpolated k-space without access to a calibration set by enforcing \textbf{P}arallel \textbf{I}maging-inspired \textbf{S}elf-\textbf{Co}nsistency (PISCO) as follows:

\begin{equation}\label{eq:PISCO_weightsolving}
   \mathrm{W}_s = \arg \min_{\mathrm{W}} \lVert \mathrm{P}_s \mathrm{W} - \mathrm{T}_s \rVert^2_2 +  \alpha \lVert \mathrm{W} \rVert^2_2 \text{\quad s.t. } \mathrm{T}_s, \mathrm{P}_s \subseteq y_s 
\end{equation}

\begin{equation}\label{eq:PISCO_condition}
\textbf{PISCO: } \sum_{i=1}^{N_s} \sum_{j=1}^{N_s} \text{distance}(\mathrm{W}_{i}, \mathrm{W}_{j}) \xrightarrow{!} 0
\end{equation}

\subsection{Dynamic MRI Reconstruction using NIK with PISCO}
\label{sec::NIK_PISCO}
In dynamic MRI reconstruction, the image reconstruction problem from Sec.~\ref{sec:methods_grappa} is expanded by a temporal dimension $t$, i.e. k-space coordinates $k$ consist of $k_i = [k_x, k_y, t]$ and result in multiple temporal images $x=\{x_t|t=1,2,...,N_t\}$. 
NIK learns a mapping $G_\theta: k \xrightarrow{} y$ based on acquired data points \cite{Huang_2023}. For abdominal reconstruction, $t$ can be derived from a self-navigator reflecting the current respiratory state \cite{Spieker_2024}. At inference, a fully-sampled spatiotemporal \mbox{k-space}~$\tilde{k}$ can be queried to reconstruct the temporal images ${x} = \mathrm{A} G_{\theta}(\Tilde{k})$.

Within each training iteration of NIK, a coordinate batch $k_B = \{{k_i}\}^{N_b}_{i=1}$ is sampled from acquired coordinates $k_{meas}$ to predict $y_B^{pred}= G_\theta(k_B)$(Fig. \ref{fig:overview}B). Then, the model is optimized with a data consistency loss $\mathcal{L}_{DC}$ comparing the predicted $y_B^{pred}$ and measured signal $y_B^{meas}$. This limits NIK's training strategy to actually acquired points. By adding the concept of PISCO during training we enable further self-supervision to coordinates that have not been acquired, which is especially beneficial for addressing large gaps in the outer regions of k-space in radial trajectories. As in Sec. \ref{sec:methods_grappa}, we assign $k_B$ as targets $k_B^T$ and sample spatial neighbors $k_B^P$. To account for the dynamic nature, we sort the coordinates along $t$ to allow for separation into $N_s$ temporally adjacent subsets~\cite{Breuer_2005} with $N_m$ coordinates each. 
The resulting $\mathrm{T}_s$ and $\mathrm{P}_s$ are 
used to solve for $[\mathrm{W}_s]_{s=1}^{N_s}$ (Eq. \ref{eq:PISCO_weightsolving}). To fulfill Eq. \ref{eq:PISCO_condition}, we introduce the PISCO loss:
\begin{equation}\label{eq:PISCO_loss}
    \mathcal{L}_{PISCO} = \frac{1}{N_s^2} \sum_{i=1}^{n_s} \sum_{j=1}^{n_s} \mathcal{L}_{dist}(\mathrm{W}_{i}, \mathrm{W}_{j})
\end{equation}
where $\mathcal{L}_{dist}$ is chosen to be the complex difference $\mathcal{L}_{\mathbbm{C}}^1 = ||\mathit{Re}(W_i - W_j)||_1 + ||\mathit{Im}(W_i - W_j)||_1$.
After pre-training NIK for $E_{pre}$ epochs, the self-supervised consistency is enforced by optimizing the model's weights $\theta$ with $\mathcal{L}_{DC}$ and $\mathcal{L}_{PISCO}$ in an alternating manner, with the latter's impact weighted by $\lambda$. Additional robustness against outliers in T and P is enhanced with Thikonov regularization weight $\alpha$ when solving for $W$. Moreover, the number of samples per equation system is increased to $N_m = f_{od} \cdot N_w$, with $f_{od}>1$ determining the ratio of unknowns $N_w$ to available equations $N_m$. The integration of $\mathcal{L}_{PISCO}$ in NIK's training procedure is summarized as pseudo-code in Alg. \ref{alg:pisco}. 

\begin{algorithm}[t]
\caption{Neural Implicit k-Space Representation with PISCO}
\begin{algorithmic}[1]
\Require Acquired k-space coordinates $k^{meas}$ and signal values $y^{meas}$, total epochs~$E$, pretraining epochs $E_{pre}$, NIK architeture $G_\theta$, initial model parameters $\theta_0$
\For{$e = 0$ to $E$}
    \For{$\text{batch ($k_B, y_B$) in } (k^{meas}, y^{meas}$)} \Comment{Sample batch of coordinates}
        \State $y_B^{pred} \gets G_{\theta_e}(k_B)$            \Comment{Predict k-space with NIK} 
        \State $\theta_{e+1}^{} \gets L_{{DC}}(y_B^{pred}, y_B^{}) $   \Comment{Update model with data consistency loss}
        \If{$e > E_{\text{pre}}$}                                            \Comment{Apply PISCO regularization}
            \State $k_B^T, k_B^T \gets k_B, \text{patches around } k_B$           \Comment{Sample coordinate patches}                     
            \State $y_B^{T}, y_B^{P} \gets G_{\theta_{e}^{}}(k_B^T), G_{\theta_{e}^{}}(k_B^P)$  
            \State $[P_s]_{s=1}^{N_s}, [T_s]_{s=1}^{N_s} \gets$ divide ${y_B^{T}, y_B^{P}}$ into multiple subsets   \Comment{Sec. \ref{sec:methods_grappa2pisco}}                     
            \State $[W_s]_{s=1}^{N_s} \gets$ solve each subset for weights with regularization $\alpha$    \Comment{Eq. \ref{eq:PISCO_weightsolving}}
            \State $\theta_{e+1}^{} \gets \lambda \cdot L_{{PISCO}}([W_s]_{s=1}^{N_s}) $      \Comment{Eq. \ref{eq:PISCO_loss}}
        \EndIf     \EndFor
\EndFor
\State \textbf{return} Learned NIK model $G_\theta$
\end{algorithmic}
\label{alg:pisco}
\end{algorithm}

\section{Experimental Setup}
\subsection{Data}
Due to the lack of ground-truth data for motion-resolved imaging \cite{Spieker_2023}, we create a dynamic free-breathing \textit{simulation} using the XCAT phantom \cite{Segars_2010}. We generate water/fat/susceptibility maps \cite{Collins_2002,Maril_2005} for 100 time points $t_{MS}$ within one breathing cycle and simulate complex images $\textbf{x}_{t}$ based on a water-fat model \cite{Yu_2008} (echo time $T_e$=1.4ms). After applying six simulated coil sensitivity maps, radial motion-free multi-coil k-space data $Y_{mf}$ is obtained using NUFFT for acceleration factors R=1,2,3. 
Then, $Y_{mf}$ is merged into one motion-affected k-space $Y_{ma}$ based on a respiratory navigator $t$ extracted from the lung-liver edge. 

For in-vivo validation, datasets were acquired using a pseudo golden angle stack-of-star trajectory (voxel size=1.5x1.5x3mm³, 
flip angle=10°) at 3T (Ingenia Elition X, Philips Healthcare) after local ethics committee approval.
We obtained a \textit{dynamic} free-breathing dataset of the abdomen, as well as a respiratory-gated acquisition from the same subject as reference. To validate the regularization independent of motion, a \textit{static} dataset of the thigh region was acquired and retrospectively accelerated (R=1,2,3) 
. Each k-space consists of 26 coils ($N_{C}$), 536 frequency encoding steps ($N_{FE}$), 1341/537/537 spokes ($N_{PE}$) for dynamic/gated/static. Coil sensitivities are estimated with ESPIRiT \cite{Uecker_2014}. 

\subsection{Training and Inference}
We adapt NIK's \cite{Huang_2023} architecture using 4 layers, 512 hidden features, high-dynamic range loss as $\mathcal{L_{DC}}$, SIREN activations \cite{Sitzmann_2020}, batch size of 10k and use STIFF feature encoding \cite{Catalan_2023}. 
We correct the respiratory signal $t$ for linear drifts and rescale it to [0,0.5]. For PISCO-NIK, $L_{\text{PISCO}}$ is included after $E_{pre}=200$ and weighted to match the magnitude of $L_{\text{DC}}$, i.e., $\lambda$=0.01/0.1 for simulation/in-vivo data. For the neighbors P, a kernel of size [3,3] with $\delta x / \delta y = N_{FE}^{-1}$ is sampled around the target T resulting in $N_n$ = 8, as shown in Fig. \ref{fig:overview}. Hence, $N_w$ for the phantom is 288 $(8 \cdot 6 \cdot 6)$ and for the subject would be 5408 $(8 \cdot 26 \cdot 26)$. To increase the number of possible sets for the subject ($N_s \propto N_w^{-1}$), we reduce the number of output coils solved for in each iteration, thereby lowering $N_w$ to 624 $(8 \cdot 26 \cdot 3)$. 
Weight solving for $L_{\text{PISCO}}$ is regularized with $\alpha$=1e-4 and $f_{OD}$=1.1. 
Both losses are optimized with separate Adam optimizers (lr=3e-5). All models are run for 
1000 epochs total (NVIDIA RTX A6000, using Python 3.10.1/PyTorch 1.13.1).


\subsection{Evaluation}
Dynamic reconstructions using $L_{\text{PISCO}}$ (PISCO-NIK) are compared to vanilla NIK (same hyperparameters, only $L_{\text{DC}}$), inverse NUFFT for 1 or 4MS (INUFFT/INUFFT4) and state-of-the-art motion-resolved XD-GRASP \cite{Feng_2016} 
using the standard 4 MS as well as 50 MS to match the temporal resolution of NIK (XD-GRASP4 and XD-GRASP50). Quantitative metrics are computed in reference to the motion-resolved $k_{mf}$ (simulation) and INUFFT-R1 (static). Calculated metrics include peak signal-to-noise ratio (PSNR) and spatial feature similarity (FSIM) of $xy$-reconstructions at 50 time points. To evaluate temporal performance, FSIM-t is calculated for the simulation on $xt/yt$ with $y/x$ fixed.

\section{Results}

\begin{figure}[b]
\includegraphics[width=\textwidth]{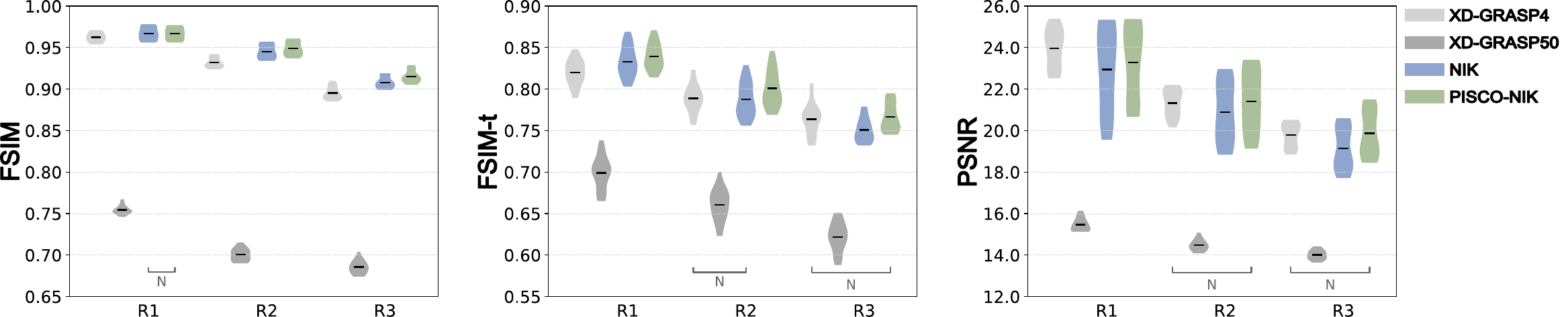}
\caption{Quantitative results for 20 motion-affected simulation slices accelerated by R=1,2,3. All comparisons, except those marked with "N", are statistically significant (Wilcoxon signed rank test with False Discovery Rate correction at p$<$0.05).} \label{quant_results}
\label{fig:quant_results}
\end{figure}
\begin{figure}[t]
\centering
\includegraphics[width=0.6\textwidth]{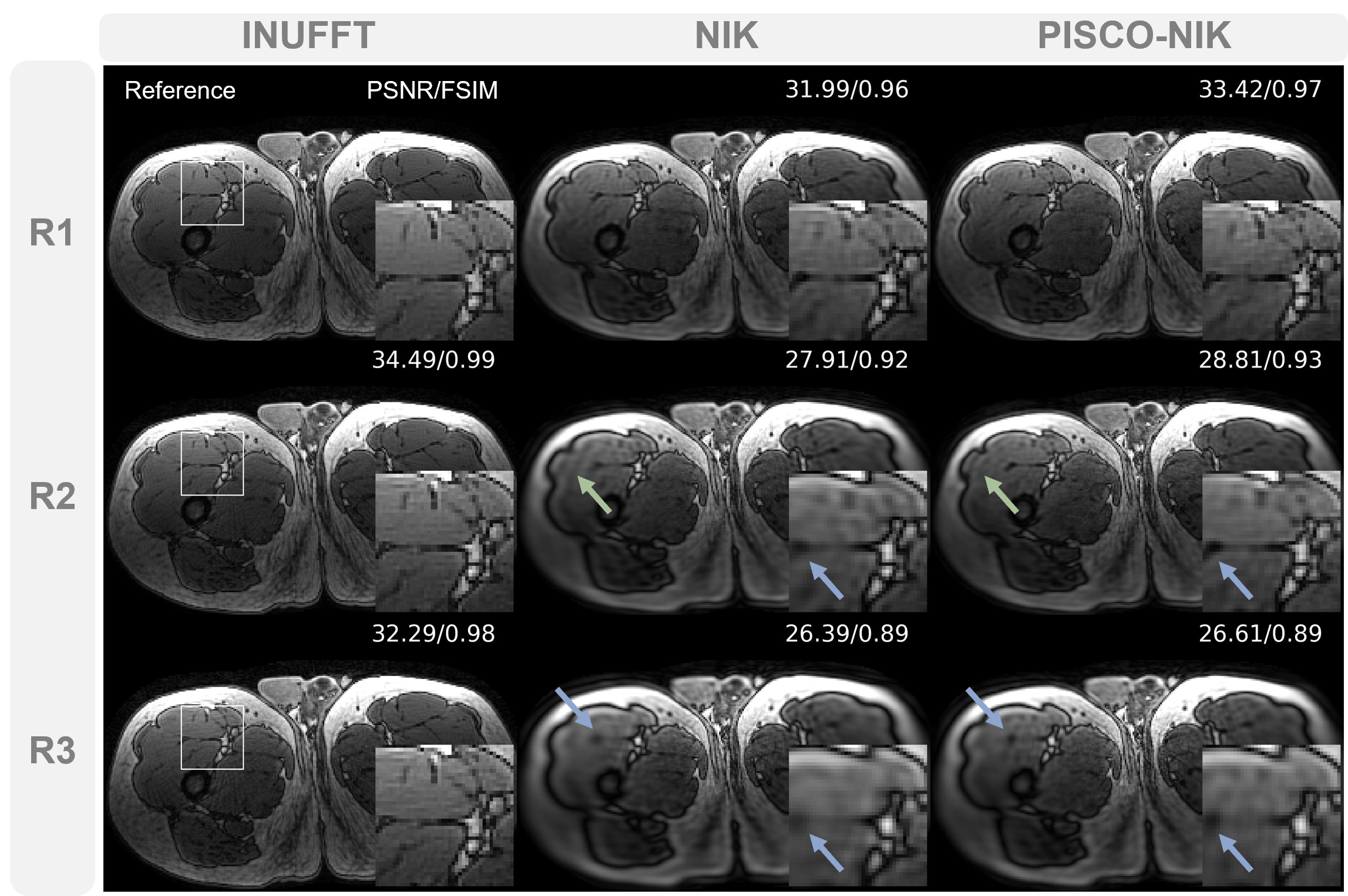}
\caption{\textit{Static} in-vivo reconstructions (thighs) for accelerations R=1,2,3. 
Exclusion of the temporal component enables comparison to INUFFT-R1, showing increased PSNR using PISCO regularization. PISCO-NIK sharpens reconstructions (blue arrows and increased FSIM) and reduces ringing artefacts of NIK-R2 (green arrows). } 
\label{fig:static_results}
\end{figure}
Quantitative reconstruction results of 20 slices of the simulation for R=1,2,3 are shown in Fig. \ref{fig:quant_results}. Videos of dynamic reconstructions are included in Suppl. Fig. 1. Both neural implicit representations, NIK and PISCO-NIK, clearly outperform XD-GRASP50 (i.e. with same temporal resolution). XD-GRASP4 results in higher PSNR due to lower temporal resolution, but significantly lower spatial FSIM due to residual blurring. 
Particularly with increasing acceleration (R=2/R=3), PISCO-NIK results in an significant increase of PSNR (up to 1.1dB), FSIM and FSIM-t (up to 0.01/0.02).
\begin{figure}[h!]
\includegraphics[width=\textwidth]{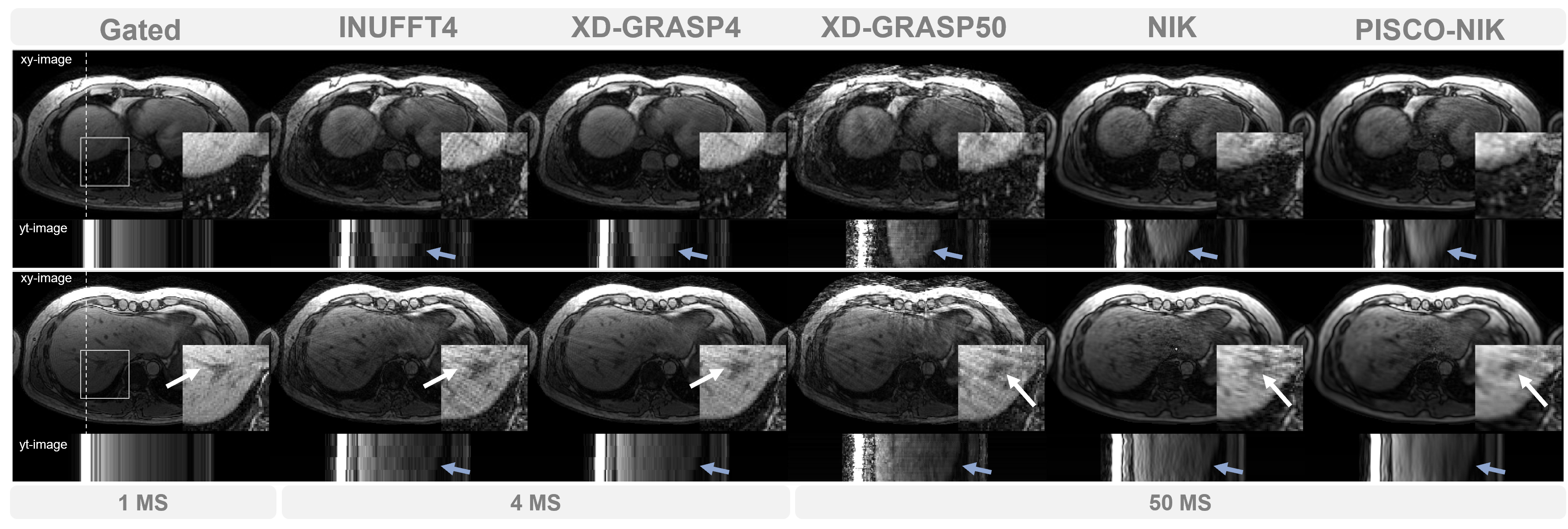}
\caption{\textit{Dynamic} in-vivo reconstructions for two abdominal slices, with spatial $xy$-image (fixed $t$) and temporal $yt$-image (fixed $x$ at white dotted line). Complete reconstruction videos in Suppl. Fig. 2.
Temporally resolved reconstructions are not available for the gated reference and 
limited for INUFFT4/XD-GRASP4 (blue arrows). XD-GRASP50 results in streaking, but more precise vessel structure, likely due to less motion blurring (white arrow). NIK enables improved temporal resolution, whereas PISCO additionally temporally smoothens (blue arrow) while maintaining spatial sharpness (white arrow). 
}

\label{fig:dynamic_results}
\end{figure}

Static reconstruction results are shown in Fig. \ref{fig:static_results}. Similar to the simulation, including PISCO results in higher PSNR and leads to sharper results, specifically with increasing acceleration factors. Note that for the static case all data are considered as one MS, therefore naturally resulting in lower undersampling for INUFFT compared to the dynamic case with multiple MS.
In Fig. \ref{fig:dynamic_results}, motion-resolved reconstructions of the dynamic in-vivo data demonstrate the benefit of neural implicit representations regarding temporal resolution ($yt$ images of NIK and PISCO-NIK, also visible in videos in Suppl. Fig. 2). Since no ground truth reconstruction is available for dynamic imaging, an unpaired gated scan is included as comparison. It results in sharp spatial images, but no temporal resolution. XD-GRASP4 smoothens results compared to INUFFT4, but residual motion blurring and discrete jumps in the temporal dimension persist (blue arrow). XD-GRASP50 encounters these visible jumps, but results in streaking. NIK allows for high temporally resolved but noisy reconstructions and the proposed PISCO regularization enforces smoother results with sharper vessel structures. 

\section{Discussion and Conclusion}
In this work, we have presented the novel concept of parallel imaging-inspired self-consistency (PISCO), which leverages global k-space relationship without the use of calibration data to explicitly determine it. We have shown the seamless integration of PISCO as self-supervised k-space regularization in NIK-based reconstructions, eliminating the need for additional data or training steps. 
Both simulated and in-vivo dynamic reconstruction results illustrate that PISCO enhances spatial and temporal denoising while preserving temporal resolution and sharp vessel structures. Additionally, we have verified PISCO's spatial denoising potential on a static example, highlighting its capacity to learn improved neural representations resulting in enhanced image quality.

Generally, NIK's flexibility in sampling and PISCO's calibration-free nature allow for a flexible kernel design. 
This is particularly beneficial for radial acquisitions, where GRAPPA calibration with a radial kernel would demand multiple ACS acquisitions due to the limitation of neighborhood relationships to small radial sections \cite{Seiberlich_2011}. In contrast, our work illustrates how the proposed PISCO-NIK approach effectively utilizes Cartesian relationships to refine radial acquisitions.


Adaptation of kernel design to the undersampling pattern may further enhance the effectiveness of PISCO. Notably, improved performance can be observed for accelerated reconstructions, particularly for R=2. This improvement can be attributed to the kernel design used, which samples one adjacent neighbor in each direction to the available acquisition points. Thereby, it effectively covers points not present with R=2. For higher acceleration factors, larger k-space gaps persist, potentially requiring a larger kernel size to focus PISCO's attention during model training at these gaps. Kernel design as well as hyperparameter tuning for PISCO are subject to further investigation. 


By introducing a regularization term for neural implicit k-space representations for the first time, we have improved its potential for motion-resolved abdominal imaging. Yet, training times and the need for re-training of neural implicit \mbox{k-space} representations remain a challenge for MR reconstruction \cite{Spieker_2023}. Moreover, the reliability of the learned representation depends on a reliable surrogate motion signal, introducing uncertainties. Future research addresses these uncertainties as well as methods to accelerate the training procedure.

Concluding, we have demonstrated how a conventional parallel imaging concept can be leveraged as self-supervised regularization for learning-based reconstruction. Due to its calibration-free and flexible design, our proposed method PISCO can be seamlessly integrated into the clinical workflow, making it an attractive regularization method for application in further anatomies or k-space-based reconstruction methods.

\bibliographystyle{splncs04}
\newpage
\bibliography{main}
%




\end{document}